\documentclass[11pt]{article}
\usepackage{appb,epsfig}
\def\la{\langle} 
\def\ra{\rangle} 
\def\be{\begin{eqnarray}} 
\def\ee{\end{eqnarray}}

\newcommand{\eq}{\begin{equation}} \newcommand{\eqx}{\end{equation}}
\newcommand{\dl}{\delta}
\newcommand{\eqn}{\begin{eqnarray}} \newcommand{\eqnx}{\end{eqnarray}}
\newcommand{\f}[2]{\frac{#1}{#2}}

\newcommand{\Tr}{\mbox{\rm Tr}}

\newcommand{\Sg}{\Sigma}
\newcommand{\cor}[1]{\left\langle{#1}\right\rangle}

\newcommand{\lm}{\lambda}

\newcommand{\inm}{I_{N_{\!f}\!-\!1}}
\newcommand{\inp}{I_{N_{\!f}\!+\!1}}
\newcommand{\knm}{K_{N_{\!f}\!-\!1}}
\newcommand{\knp}{K_{N_{\!f}\!+\!1}}

\newcommand{\uone}{$U(1)$ {}}

\newcommand{\tr}{\mbox{\rm Tr}}
\newcommand{\gm}{\gamma}

\begin{document}
\title{Chiral Random Matrix Models in QCD}
\author{Romuald A. JANIK\address{Service de Physique Th\'{e}orique, 
  CEA Saclay, F-91191, Gif-sur-Yvette, France;\\
  Department of Physics, Jagellonian University, 30-059, Krakow,
  Poland.}
\and Maciej A. NOWAK\address{Department of Physics, Jagellonian
  University, 30-059, Krakow, Poland;\\
  GSI, Planckstr. 1, D-64291 Darmstadt, Germany.}
\and G\'{a}bor PAPP\address{CNR Department of Physics, KSU, Kent, 
  Ohio 44242, USA;\\ 
  Institute for Theoretical Physics, E\"{o}tv\"{o}s University,
  Budapest, Hungary.}
\and Ismail ZAHED\address{Department of Physics and Astronomy,
  SUNY, Stony Brook, N.Y. 11794, USA.}
}
\maketitle



\begin{abstract}
We review some motivation behind the introduction of chiral random matrix
models in QCD, with particular emphasis on the importance of the 
Gell-Mann-Oakes (GOR) relation for these arguments. We show why the 
microscopic limit is universal in power counting, and present arguments 
for why the macroscopic limit is generic for a class of problems that defy 
power counting, examples being the strong CP and $U(1)$ problems. Some
new results are discussed in light of recent lattice simulations.
\end{abstract}

\section{Introduction}

It is well known that the spontaneous breaking of chiral symmetry in QCD
along with its explicit breaking, provide stringent constraints on
on-shell amplitudes~\cite{WEINBERG}, whether at threshold or
beyond. These constraints can be exploited in power counting
(ChPT)~\cite{LEUTWYLER}, or to any order~\cite{YAZA}, leading to a
comprehensive description of a number of reaction processes. The same
symmetry constrains the off-shell amplitudes such as the vacuum to
vacuum amplitude in the presence of external sources.

What is less well-known perhaps, is the fact that in a finite Euclidean
volume $V=L^4$ similar constraints are still in action even in the case 
where the box size $L$ is smaller than the pion Compton
wavelength $1/\sqrt{m}$ (in units where the QCD scale $\Lambda=1$).
This is a regime where the spontaneous breaking of chiral symmetry is 
obsolete. This observation for the off-shell amplitudes was first
emphasized by Gasser and Leutwyler~\cite{GASSER}. Their observation
carries to most models with spontaneously broken symmetries in
$d>2$ as was discussed by Hasenfratz and Leutwyler~\cite{HASENFRATZ}.

The regime for which $mV>1$ will be referred to as macroscopic. In it,
chiral symmetry is spontaneously broken with $\la q^{\dagger} q\ra=1$
(again in units where $\Lambda=1$). The regime for which $mV<1$ will be
referred to as microscopic. In it, chiral symmetry is effectively
restored with $\la q^{\dagger} q\ra= mV +{\cal O}$. The nature of the
corrections ${\cal O}$ will be discussed below. The regime $mV\sim 1$
will be referred as transitional.  In it, chiral symmetry is about to be
restored or broken depending on how we wish to look at it. Throughout,
the box size is chosen $L>1/{\Lambda}=1$, so that nonperturbative
effects are present. At small but finite temperature, a similar classification
holds in Euclidean space with $V=\beta L^3$.
\begin{figure}[htbp]
\centerline{\epsfysize=30mm \epsfbox{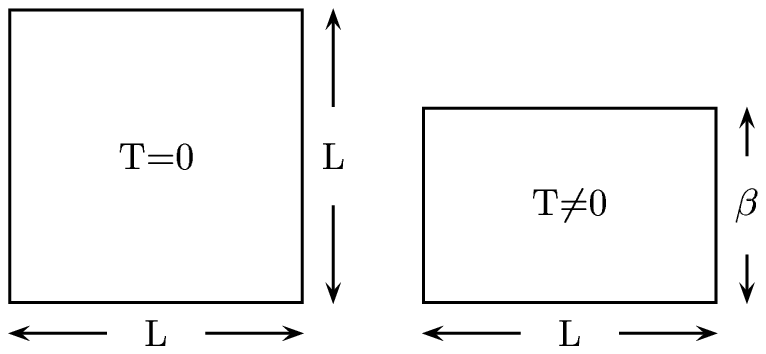}}
%
\caption{QCD in a symmetric (left) and asymmetric (right) box.}
\label{figqcd}
\end{figure}

The importance of the parameter $mV$ for the constant
mode problem in QCD was first noted by Jolicoeur and Morel~\cite{MOREL}
in the strong coupling regime. This remark is important as it implies that
the observations to follow carry for both strong and weak coupling provided
that a continuous symmetry is broken spontaneously with the occurrence of
Goldstone bosons. 

In the present work, we will go over a number of observations in the
context of chiral symmetry that paves the road for the onset of chiral
random matrix models in QCD with their relevance to the three regimes of
$mV$ detailed above. In section 2, we go over the original observation
by Banks and Casher~\cite{BANKS}. In section 3, we detail some of the
arguments given by Gasser and Leutwyler~\cite{GASSER} regarding the
finite volume partition function, paying particular attention to the
Gell-Mann-Oakes-Renner (GOR) relation. In section 4, we summarize how these
observations were used by Leutwyler and Smilga~\cite{SMILGA} to derive
spectral sum rules, and derive new spectral sum rules for generalized GOR
relation. In section 5, we present simple arguments for why
these sum rules are minimally reproduced by chiral random matrix
models. In section 6, we go over the construction by Verbaarschot and
Zahed~\cite{VERBAR} for the microscopic n-point correlation functions
as the master formulas for all diagonal and off-diagonal sum rules. 
The n-point correlation functions specify the degree of flavor mixing
in the n-vacua, and uniquely condition the bulk spectral rigidity and level
variance. The relevance of these observations for lattice simulations in
symmetric and asymmetric boxes is discussed. In section 7, we discuss
the microscopic spectral density in the double scaling limit,
following the original arguments
by Jurkiewicz, Nowak and Zahed~\cite{JUREK} and more recently Damgaard
and Nishigaki~\cite{DAMGAARD} and Wilke, Guhr and
Wettig~\cite{WILKE}.  New sum rules with decoupled sea and valence quark 
effects are discussed.
We argue  that the associated spectral
density in the double scaling limit is universal in QCD, within a range
of quark masses that we discuss. In section 8, we put forward
arguments for why chiral random matrix models are of interest in the
macroscopic regime, especially when chiral power counting breaks
down. We illustrate our points for the CP and \uone problems. In section
9 we summarize our conclusions.

\section{Quark Spectrum}

In the early $80$'s Banks and Casher made a remarkable observation regarding 
the relation of the quark condensate to the QCD quark spectrum. In an Euclidean 
volume $V$, they noticed that the amount of quark pair condensation 
 $\la q^{\dagger} q\ra$ relates in a simple way to the distribution of quark 
eigenvalues $\nu (\lambda )$. Indeed, 
\be
V\langle q^{\dagger} q \rangle&& = \sum_k \langle
\frac 1{\lambda_k + im} \rangle \nonumber\\
&&=\int d\lambda \frac 1{\lambda + im } \langle {\rm Tr} \delta (\lambda 
-iD)\rangle\nonumber\\
&&=\int d\lambda \frac {-im}{\lambda^2+ m^2} \nu (\lambda)
\rightarrow -i\pi \, \nu (m)
\label{banks1}
\ee
where the limit is understood for $mV>1$. The quark eigenvalues $\lambda_k$
depends implicitly on the given gauge configuration, through the eigenvalue
equation $iD\, q_k =\lambda_k \, q_k$. They are paired, since $[\gamma_5, 
iD]_+=0$. This pairing causes $\nu (\lambda)$ to be
an even function on its support. The averaging in (\ref{banks1}) is over the 
gauge configurations.

In (\ref{banks1}) the zero virtuality point $\lambda=0$
\footnote{The quark virtuality $\lambda$ is analogous to a complex mass
in 4d Euclidean space, or  an energy in 1+4d Minkowski space.}
is the analogue of a Fermi surface. This
observation suggests in fact that the regime $mV>1$, in which chiral 
symmetry is spontaneously broken, is actually diffusive~\cite{PRLUS}, 
following the delocalization of the quark eigenmodes.

More remarkably, however, is the fact
that for the pair condensate to be non-zero, the density of quark eigenmodes
near zero virtuality has to {\it grow} with the volume $V$. In other words, 
the level spacing of the quark eigenmodes in a finite volume $V$ and near zero 
has to scale as $1/V$ in sharp contrast to the level spacing of a free particle 
in the same box which has to scale as $1/L$.

\begin{figure}
\centerline{\epsfysize=30mm \epsfbox{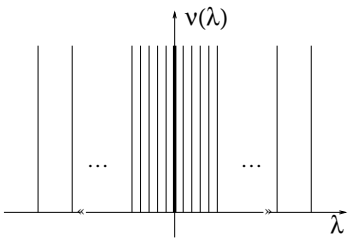}}
\caption{Quark spectrum in the  vacuum.}
\label{qcdvac}
\end{figure}

The spontaneous breaking of chiral symmetry causes a large accumulation
of quark eigenmodes near zero virtuality as pictured in
Fig.~\ref{qcdvac}, for a fixed gauge-configuration. The occurrence of
exact zero modes of the nonzero $n$ (winding number) configuration is
geometrical and follows from the 
topological character of the background gauge configuration. Conversely,
any model that breaks spontaneously the underlying chiral symmetry
displays a spectrum of the type shown in Fig.~\ref{qcdvac}. This means
QCD, instanton models, NJL models, etc.

\section {Finite Volume Partition Function}

\subsection{With GOR}

Few years later, Gasser and Leutwyler~\cite{GASSER} noticed
that in a finite volume $V$ for which the Compton wavelength of 
the pion is large, that is $mV<1$ but $L>1$, while chiral symmetry is 
effectively restored with $\la q^{\dagger} q\ra= mV +{\cal O}$, the way 
it is restored is still {uniquely} conditioned by the way chiral symmetry is 
about to be spontaneously broken. For QCD this observation is natural if we 
were to approach the microscopic regime $mV<1$ from the macroscopic regime 
$mV>1$ by fine tuning the bare quark mass $m$.

{\bf $mV>1$ region:\,\,\,}
In a finite Euclidean box $V$ the finite volume partition function $Z$
is the sum of zero-point motions of singlet excitations: pions, kaons,
nucleons, etc. (Casimir effect). In the regime $mV>1$ the spontaneous
breaking of chiral symmetry together with the Gell-Mann-Oakes-Renner
(GOR) relation~\cite{GOR} allows for a book-keeping of ${\rm ln} Z$ by
using chiral power counting with a typical momentum scale $p$, where
 $U\sim p^0$, $1/L\sim p$ and $m \sim m_\pi^2\sim p^2$ and $m^2 V\sim 1$
fixed. In this book-keeping,
the contribution to order $p^2$ is given by the following Lagrangian
\cite{GASSER}

\be
{\cal L}_2 = \frac {F^2}4 {\rm Tr} \left(\partial_{\mu} U^{\dagger} 
\partial_{\mu} U\right)
 -  \frac 12\Sigma \,\,{\rm Tr} ( m U + 
U^{\dagger}m^{\dagger})
\label{gas1}
\ee
for $N_f$ flavors with $U=e^{i\hat{\pi}/F}$ 
($\hat{\pi}=\pi^\alpha \lambda_\alpha$) an $SU(N_f)$ valued field subject
to the boundary condition $U(x+L)=U(x)$. The normalization is chosen
$\mbox{Tr} \lambda_{\alpha}\lambda_{\beta}= 2\delta_{\alpha,\beta}$.
Throughout we will use $m$
for either the non-degenerate case $m=(m_1, ..., m_{N_f})$ or the degenerate
case $m=m{\bf 1}_{N_f}$. We hope that the reference to each case will be clear 
from the notation used. 

In a periodic box, the free pion propagator is
\be
G_{\pi} (x) = \frac 1V \sum_{k_n} \frac{e^{-ik_n x}}{k_n^2 + m_{\pi}^2} =
\frac 1{Vm_{\pi}^2} + \frac 1V \sum_{k_n}' 
	\frac{e^{-ik_n x}}{k_n^2 + m_{\pi}^2}
\label{gas2}
\ee
where the primed sum is over non-zero four momenta with $k_n = (n_1,n_2,n_3, 
n_4 )$ in units of $2\pi/L$. To order $p^2$ the pion mass is
 $m_{\pi}^2=m \Sigma/F^2 \sim m$ and the condensate is 
 $-\la\bar{q}q\ra=\Sigma$. To order $p^4$, the one-loop
contribution from  
(\ref{gas1}) has to be supplemented with the most general terms consistent 
with broken chiral symmetry and general principles~\cite{GASSER}
(Lorentz invariance is upset in a box to order $p^4$). 
\begin{figure}[htbp]
\centerline{\epsfysize=15mm \epsfbox{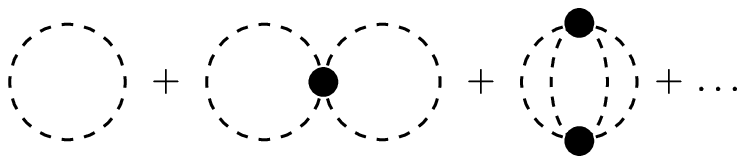}}
\caption{Meson contribution to ${\rm ln} Z$ for $mV>1$.}
\label{fig-mesg}
\end{figure}

{\bf $mV<1$ region:\,\,\,}
In the microscopic regime $1>mV$, the zero mode contribution
$1/mV$ in (\ref{gas2}) dwarfs the
non-zero mode contributions (primed sum) causing 
the chiral power counting to break down. In this regime Gasser and Leutwyler
suggested a reorganization of the chiral power counting fixing $mV$
instead of $m^2V$ with expansion in $\epsilon=1/L$, thereby allowing for
a systematic resummation  
of the pionic zero modes in the finite volume partition function. 

To order ${\cal O}(\epsilon^2)$ the dependence of $Z[S,P,V]$ on the
external sources $S=m\Sigma/2 +S'$ and $P$ is simply
\be
Z[S,P,V] = N[0] \int_{SU(N_f)}\hspace*{-8mm}
	 dU\,e^{V {\rm Tr} ((S+iP) U+{\rm h.c.})}
	\,.
\label{gas3}
\ee
$N[0]$ receives contribution from the non-zero mode terms, essentially the 
Casimir energy of free pions in a box V as indicated by the first diagram
in Fig. 3. This contribution is $S,P$-independent to the order quoted. For
$x=m\Sigma V$ and 
$N_f=1$, $Z [m]\sim e^{x}$, while for $N_f=2$ with equal masses it is
$Z[m]\sim I_1 (2x)/2x$~\cite{GASSER}. For one massive flavor and
$N_f-1$ massless flavors it is $Z[m]\sim I_{N_f-1} (x)/x^{N_f-1}$.
In each case, $Z[m]=Z[m,0,V]$ and the proportionality constant is $m$ 
independent. 

{}From (\ref{gas3}) 
the chiral condensate to order ${\cal O}(\epsilon^2 )$ is 
\be
i\la q^{\dagger} q\ra= -\frac {N[0]}{Z[m]} \frac{\Sigma}2
\int_{SU(N_f)}\hspace*{-8mm} dU\, {\rm Tr} (U+U^{\dagger})
e^{\frac 12 m\Sigma V {\rm Tr} (U+U^{\dagger})}
\label{gas4}
\ee
and vanishes as $x=mV\rightarrow 0$, to the exception of $N_f=1$
where $i\la q^{\dagger} q\ra =-\Sigma$. For $N_f=2$ and two equal
masses, $i\la q^{\dagger} q\ra =-\Sigma I_2 (2x)/I_1 (2x)$, while
for one massive and $N_f-1$ massless quarks 
 $i\la q^{\dagger} q\ra =-\sigma I_{N_f} (x)/ I_{N_f-1} (x)$. 
In this limit all points on
$SU(N_f)$ are equally weighted. Although chiral symmetry is restored in
the regime $mV<1$, the way by which it is restored is conditioned
by :
\vskip .5cm
\noindent {\bf i)} the nature of the coset;
\newline {\bf ii)} the explicit $(N_f,N_f)$ breaking; 
\newline {\bf iii)} the underlying GOR assumption. 
\vskip .5cm


\begin{figure}[htbp]
\centerline{\epsfysize=15mm \epsfbox{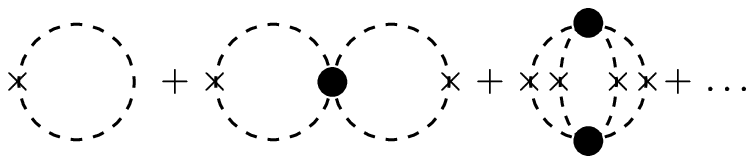}}
\caption{Zero momentum contribution to ${\rm ln} Z$ for $mV<1$.}
\label{fig-mess}
\end{figure}

\subsection{Without GOR}

{\bf $mV>1$ region:\,\,\,}
The GOR assumption is important in establishing (\ref{gas1}) as the
starting point for power counting in ChPT in the macroscopic regime. 
This assumption may be relaxed in the form of generalized ChPT. An 
example to order $p^2$ is
\be
{\cal L}_2 =&& \frac {F^2}4 {\rm Tr} \left(\partial_{\mu} U^{\dagger} 
\partial_{\mu} U\right)
 - \frac 12 V\Sigma \,\,{\rm Tr} ( m U + U^{\dagger}m^{\dagger})\nonumber\\
&& + A {\rm Tr} (m^{\dagger} U m^{\dagger} U +{\rm h.c.}) 
+ B \left({\rm Tr} (m^{\dagger} U +{\rm h.c.} )\right)^2 \nonumber\\
&& + C\left( {\rm Tr} (m^{\dagger} U - {\rm h.c.} )\right)^2 + 
D{\rm Tr} (m^{\dagger} m)
\label{gas5}
\ee
as suggested by Stern, Sazdjian and Fuchs and others~\cite{general}. 
Here $\Sigma$ and $m$ are 
counted of order $p$, while $A,B,C,D$ are counted of order $p^0$.
The constants $\Sigma, A,B,C$ are low-energy parameters, while $D$ is not
(see below).  To order $p^2$ the pion mass 
squared and the chiral condensate are
\be
m_{\pi}^2 = &&\frac{m}{F^2} \left(\Sigma - 8m A - 8m N_fB\right)\nonumber\\
-\langle\overline{q} q\rangle = &&\Sigma - m(2 A+D + 4N_fB) \,.
\label{gas55}
\ee
The constant $D$ is subtraction dependent. It relates to the 
$m$-dependent quadratic divergence in the condensate. It is not
amenable to low-energy constraints. 

{\bf $mV<1$ region:\,\,\,}
Since the contribution of the pionic zero modes in (\ref{gas2}) are of the 
form $1/V{m_{\pi}}^2$, the finite volume partition function (\ref{gas3}) for 
fixed $Vm_{\pi}^2$ would involve $A,B,C,D$ from (\ref{gas5}). The microscopic 
limit is changed compared to the GOR assumption if we were to take
 $Vm_{\pi}^2$ fixed but small (see section IVB).
Such a limit could be reached by varying $m$ for fixed $m^2 V <1$. The power
counting in $\epsilon=1/L$ would be: $\Sigma$ and $m$ of order $\epsilon^2$,
 and $A,B,C,D$ of order $\epsilon^0$. In this
regime, the chiral condensate again vanishes in the chiral limit as all 
points on the $SU(N_f)$ manifold are equally weighted.

Clearly $mV$ fixed selects the GOR form in the microscopic regime, and could 
be used for a precise determination of $\Sigma$ by comparison to ChPT. This
method may be more accurate than a simple extrapolation of the condensate to
zero $m$, owing to finite size effects. This is more than an academic 
exercise since both (\ref{gas1}) and (\ref{gas5}) lead to the same low
energy constraints, and experiments are underway at DAPHNE to test the validity
of the GOR hypothesis by firming up the accuracy of the S-wave $\pi\pi$ 
scattering length near threshold~\cite{FRANZINI}.

\section{Microscopic Sum Rules}

\subsection{With GOR}

In the early $90$'s Leutwyler and Smilga~\cite{SMILGA}
remarked that the finite volume
partition function could be used to derive sum rules for the quark
spectrum (qualitatively shown in Fig.~\ref{qcdvac}) in the microscopic
regime for gauge configurations of fixed winding number $n$. Assuming 
the concept of n-vacua for the QCD state, the finite volume 
partition function for fixed $n$ relates to the finite volume 
partition function through 
\be
Z_n [m] &=&\int_0^{2\pi}\! \frac {d\theta}{2\pi}\ e^{-in\theta} \,
Z[me^{i\theta/N_f} ]\nonumber\\
&=& N(0) \int_{U(N_f )}\hspace*{-7mm} dU\ ({\rm det}U)^n\, 
	e^{\frac 12 V\Sigma \Tr(mU +U^{\dagger} m^{\dagger})}
\label{smi1}
\ee
in the microscopic regime $mV<1$. 
For QCD $n=\frac{\alpha_s}{8\pi}\int E^a\cdot B^a$ and is valued 
in ${\bf Z}$. The left-hand side in (\ref{smi1}) involves the quark 
spectrum averaged over gauge configurations with fixed winding number $n$, 
while the right-hand side involves an information from the hadrons. 
$Z_n [m]$ is local in the quark variables, but non-local in the hadronic
variables. Scalar n-point functions following from (\ref{smi1}) do not
obey the cluster decomposition.

$Z_n [m]$ does not discriminate between
$N_f=1$ and $N_f>1$ in QCD. Also, $Z_n [m]$ is insensitive to the nature
of the sources $S'$ and $P$ in (\ref{gas3}), and only depends on the quark 
mass matrix chosen diagonal, and its overall phase for $n\neq 0$, since
\cite{SMILGA}
\be
Z_n [{\bf V} m {\bf W} ] = ({\rm det} ({\bf V W}))^n Z_n[ m] \,.
\label{smi11}
\ee
This is not the case for (\ref{gas3}).

For one flavor, we have
\be
Z_n[m]=m^{|n|} \langle \prod_{\lambda_k>0}' (\lambda_k^2 + m^2) \ra_{n,A} =
n! \,{\rm I_n} (x) \,.
\label{smi2}
\ee
with $x=mV\Sigma$ and ${\rm I}_n$ a Bessel function (the primed product
does not contain the exact zero modes and the eigenvalues with negative
real part). The averaging is
over the gauge configuration $A$ with fixed winding number $n$. Using
the infinite representation for the Bessel function and rearranging
(\ref{smi2}) it follows that
\be
\Big\la\Big\la\prod_k' \left(1+\frac{m^2}{\lambda_k^2}\right)
	\Big\ra\Big\ra_n
=\prod_k \left(1+\frac {x^2}{\xi_{n,k}^2}\right)
\label{smi3}
\ee
where $\xi_{n,k}$ are the zeros of the Bessel function $J_n (\xi_{n,k})=0$ 
with $k=1,2, ...$. The double averaging on the left-hand side of (\ref{smi3})
now includes $m^{|n|}\prod_k' \lambda_k^2$ as part of the measure,
with a normalization to $1$ at $n=0$ and $m=0$. By identifying
the same powers of $m$ on both sides of (\ref{smi3}) sum rules for the quark 
eigenvalues at fixed $n$ follow, in the microscopic regime. An example is
\cite{SMILGA}
\be
\frac 1{V^2}
\langle\langle\sum_k'\frac 1{\lambda_k^2}\rangle\rangle_n = \frac {\Sigma^2}
{4(n+N_f)}
\label{smi4}
\ee
where we have again reinstated $\Sigma$ and $N_f$ for convenience. The
contribution of 
the ultraviolet modes to the primed sum in (\ref{smi4}) vanishes like $V/V^2$
in the large volume limit.

It can be checked that the sum rule (\ref{smi4})
is just a suitable arrangement of moments directly amenable to the finite
volume partition function (\ref{gas3}). They reflect therefore on the sum
rules satisfied by the invariant QCD correlators at zero momentum in the
microscopic limit. For example 
\be
\frac 1{V^2}
\langle\langle\sum_k'\frac 1{\lambda_k^2}\rangle\rangle_n = 
\frac {\Sigma^2}{2N_f}\int_0^{2\pi}\f{d\theta}{2\pi}\,e^{in\theta}
\langle\left({\rm Tr} (e^{\frac{i\theta}{N_f}} U^{\dagger} + U
e^{\frac{-i\theta}{N_f}}) \right)^2\rangle_{\rm con}
\label{smi44}
\ee
where the averaging on the rhs is done with the Haar measure normalized
to 1. Notice that while the lhs is local in the
quark variables, it is not on the rhs, owing to the $\theta$ integration.
The rhs is the zero momentum and $\theta$-averaged contribution from the
quark susceptibility $\la \theta | (q^{\dagger} q)^2 |\theta \ra$, evaluated
in the microscopic limit $mV \rightarrow 0$, with $\la q^{\dagger}q\ra =0$.
The susceptibility reduces to the variance of the scalar operator on the
invariant measure.

The remarkable property of (\ref{smi3}) is that it implies that the quark 
spectrum near zero virtuality is indeed $1/V$ spaced, with a spacing that 
follows as if from a `master' gauge field in the form $\xi_{n,k}/V$. In
the microscopic regime the continuum limit $V\rightarrow\infty$ is taken by 
keeping the level spacing finite. The master formula for all diagonal
microscopic sum rules is just given by
\be
\nu_n (s =\lambda V) = \frac 1V \sum_{k=1}^{\infty} \delta (\lambda 
-\frac {\xi_{n,k}}V )
=\sum_{k=1}^{\infty} \delta (s - {\xi_{n,k}}) \,.
\label{smi5}
\ee
The limit $V\rightarrow\infty$ with $s$ fixed was originally discussed
by Shuryak and Verbaarschot~\cite{SHURYAK}, and the microscopic sum rules 
were numerically checked to hold for the instanton liquid model. This is
expected, since condition {\bf i)-iii)} at the end of section {\bf IIIA} 
hold verbatim for the instanton liquid model
as shown by Alkofer, Nowak, Verbaarschot and Zahed~\cite{ALKOFER}.

\subsection{Without GOR}

We note that the present results were derived starting from the GOR 
assumption with the $\epsilon$ expansion carried in the regime $mV$
fixed but small. If the GOR relation is given up as discussed in section 
{\bf III}, then in the regime $m_{\pi}^2 V\sim m^2V$ fixed the microscopic 
sum rules are changed. In particular (\ref{smi4}) becomes for $n=0$

\be
&&\frac 1{V^2}
\langle\langle\sum_k'\frac 1{\lambda_k^2}\rangle\rangle_0 = 
\left[\frac 14\Sg^2-\f{2}{V}(B-C)\right]
\frac{\cor{|\Tr U|^2}_0}{N_f}+\f{D}{V}
\label{GORREL}
\ee
The averaging on the rhs is done over $U(N_f)$ with the Haar measure. 
The $1/V$ term is of order $\epsilon^4$, but so is $\Sigma^2$.
We note, that contrary to $m$ which is a parameter, $\Sigma$ is fixed by
the QCD dynamics. In small volumes $V$ with $m_{\pi}^2 V$ held fixed but small,
the power counting is valid, and the second term 
may not be negligible. The $D$ dependence in (\ref{GORREL})
is expected from the subtraction dependence to this order.

\section{Chiral random matrix models}

With the renewed interest in the instanton liquid model along the lines 
discussed by Diakonov and Petrov~\cite{DIAKONOV} and also 
Shuryak~\cite{SHUINST}, it became clear that the bulk issues elegantly 
revealed by the variational and statistical approach to the instanton 
physics are generic and likely amenable to a description in
terms of random matrix theory~\cite{MEHTA}. This observation
is particularly relevant for the quark eigenvalues and their correlations 
as originally suggested by Nowak, Verbaarschot and Zahed~\cite{NOWAK}. 
In this context, the relevance of the three Wigner ensembles~\cite{MEHTA}
was noted earlier by Simonov~\cite{SIMONOV}, and the relation to the 
microscopic limit made later by Shuryak and Verbaarschot~\cite{SHURYAK}.
In this limit the physics is that of a rotor plus a potential
for the uncolored Goldstone 
constant modes~\cite{LEUTWYLER}, and that of random matrix theory for the 
colored constant modes.

\subsection{With GOR}

The universality of the finite volume partition function in the microscopic 
regime implies that any model that satisfies the conditions 
{\bf i)-iii)} summarized at the 
end of section {\bf IIIA}, will do as far as the microscopic sum rules are 
concerned, thanks to the observation made by
Gasser and Leutwyler~\cite{GASSER}. The minimal model that satisfy these 
requirements is a four-fermi model in 0-dimension (matrix model). Specifically,
\be
{\cal L} = q^{\dagger}\,\, im e^{i\gamma_5\theta/N_f}\,\, q
+ \frac 1{V\Sigma^2} q_R^{\dagger}q_L q_{R}^{\dagger}q_L + ...
\label{rmm1}
\ee
with ${\rm dim} q_{L,R}=N_f\times n_{L,R}$ are Grassmann numbers, 
$\gamma_5 = ({\bf 1}_R, {\bf 1}_L )$, and $n_L+n_R=V$. The $...$ 
terms in (\ref{rmm1}) could be either singlets
$(1,1)$, or non-singlets. The latters should vanish in the chiral 
limit as $m^a$ with $a>1$.

The finite volume partition function associated to (\ref{rmm1}) may be 
either cast in the Grassmann variables, or by bosonization through
${\cal A}^{ab} = {q_L}^a {q_R^b}^{\dagger}$ which is $N_fn_L\times N_fn_R$, 
turned into an integral over bosonic matrices $A$ in the form
\be
Z[\theta, m] = \int && d{\cal A}\,e^{-V\Sigma{\rm tr} ({\cal A}^{\dagger}
 {\cal A}) + ...} \,\,\,\nonumber\\ &&{\rm det}
\left(\begin{array}{cc}
ime^{i\theta/N_f}&{\cal A}\\
{\cal A}^{\dagger}&ime^{-i\theta/N_f}
\end{array}\right) \,.
\label{rmm3}
\ee
The determinant is $V\times V$ for $N_f=1$. From our preceding arguments, 
it does not matter what $...$ is in the microscopic limit $mV<1$.
The finite volume partition 
function (\ref{rmm3}) is that of a chiral random matrix model. In QCD
the quarks are in the complex representation, so $A$ is a complex valued 
matrix. The chiral random matrix ensemble generated by the description 
(\ref{rmm3}) is ChGUE.

\subsection{Without GOR}

The minimal four-fermi model in 0-dimension that is commensurate with
(\ref{gas5}) is
\be
{\cal L} = && q^{\dagger}\,\, im e^{i\gamma_5\theta/N_f}\,\, q 
+ \f{4}{V\Sigma^2} \Biggl[\,\,q_R^{\dagger}q_L q_{R}^{\dagger}q_L \nonumber\\
&& +\f{(B+C)}{ V} \, 
\Big( (q_L^{\dagger} m^{\dagger} q_L)^2
+ (q_R^{\dagger} m q_R )^2 \Big)\nonumber\\
&& +2 \f{(B-C)}{V} (q_R^{\dagger} q_R)
 (q_L^{\dagger} mm^{\dagger} q_L) \Biggr]\nonumber\\
&&  + D {\rm Tr} (m^{\dagger} m) +...
\label{rmma}
\ee
Again, ... are either singlets or non-singlets but then of higher order in
$m$. The finite volume partition function can be bosonized using:
${\cal A} =q_L q_R^{\dagger}$, ${\cal B}^R=q_R q_R^{\dagger}$ and 
${\cal B}^L=q_L q_L^{\dagger}$, the result is a multi-random
matrix model ($N_f=1$)
\be
\left(\begin{array}{cc}
ime^{i\theta} + m e^{i\theta} {\cal B}^R & {\cal A}\\
{\cal A}^{\dagger}&ime^{-i\theta}+m e^{-i\theta} {\cal B}^L
\end{array}\right)
\ee
with width of the distribution given by
\eqn
\cor{{\cal B}^L_{ab} {\cal B}^L_{cd}} &=& \cor{{\cal B}^R_{ab} {\cal
B}^R_{cd}} = \f{1}{\sqrt{V}}\cdot \f{8(B+C)}{\Sg^2} \dl_{ad}\dl_{bc} \\
\cor{{\cal A}_{ab} {\cal A}_{cd}^\dagger}&=&\f{1}{V}\cdot\f{1}{\gm_{EFF}}
\dl_{ad} \dl_{bc}
\label{width}
\eqnx
where
\eq
\f{1}{\gm_{EFF}}=\f{4}{\Sg^2}\left(1+(B-C)\f{2m^2}{V}\right)
\eqx
The inverse bosonization of (\ref{rmma}) using 
$q^\dagger_{R,L} q_{R,L}$ as bosonic variables
can be shown to yield (\ref{gas5}) as expected. 
In this case, the different scalings in $V$ in
the width (\ref{width}) are important. The terms
of order $(\tr U)^2$ follows only after an exact
gaussian integration over ${\cal A}$ and ${\cal B}$
has been performed. We note that in the present 
formulation, (\ref{rmma}) yields $A$ in terms of
$B,C$ and $\Sigma$.

\section{Microscopic Spectral Distribution}

Since the microscopic sum rules follow from chiral random matrix models in the 
microscopic regime $mV<1$, we can then use the powerful machinery of random 
matrix theory to make precise statements on the eigenvalue distribution 
and its correlation. Without loss of generality, we can just assume
a gaussian weight in (\ref{rmm3}) and proceed. Throughout, we will discuss
the GOR case. The alternative case is straightforward to implement.

\subsection{Orthogonal Polynomial Method}

Following Verbaarschot and 
Zahed~\cite{VERBAR}, we may specialize to the symmetric case $n_L=n_R=V/2=N$
(zero winding sector), and 
use the polar decomposition of the unitary matrix $A$,
$A=U\Lambda V^{\dagger}$, where $\Lambda={\rm diag}(\lambda_1, ..., \lambda_V)$
in ${\bf R}_+$, and $U$ in $U(V)$ and $V$ in $U(V)/U(1)^V$. The division by 
$U(1)^N$ preserves the number of degrees of freedom. As a result, the angular 
and radial integration separates, the former inducing only an overall 
$m$-independent normalization. It will be ignored. With this in mind, we have
\be
Z[ N, m] = \int_0^{\infty}
\prod_{i=1}^N d\lambda_i e^{V\Sigma\lambda_i^2}
\prod_f^{N_f} 
(\lambda_i^2+m_f^2) \,\,\Delta [\lambda ]
\label{mic1}
\ee
with the Vandermonde determinant
\be
\Delta [\lambda ] =\prod_{i<j} |\lambda_i| (\lambda_i^2-\lambda_j^2)^2
\label{van}
\ee
following from the Haar measure on $SU(V)$ in the polar decomposition 
\cite{BOOKMEASURE}. The first determinant in (\ref{mic1}) stems from the 
integration over the Grassmannians. 

In random matrix theory~\cite{MEHTA}, the partition function (\ref{mic1})
has a simple form in terms of the joint eigenvalue distribution
\be
\nu(\lambda_1, ..., \lambda_V) = |\langle P_1 ...P_V | \lambda_1 
...\lambda_V\rangle |^2
\label{mic2}
\ee
where the $P_i$'s form a set of ortho-normalized polynomials of degree
 $i-1$ with  
 $1\leq i\leq V$ (with the term of degree $i$ normalized to $1$), with 
the measure 
\be
\int_0^{\infty} d\lambda^2 e^{-V\lambda^2}\prod_{f=1}^{N_f}
(\lambda^2+m_f^2) P_i(\lambda )P_j(\lambda ) = {\bf 1}_{ij} \,.
\label{mic3}
\ee
The relation (\ref{mic2}) follows simply from the properties of determinants
(invariance under arbitrary linear combinations of either columns or rows). 
It is a Slater determinant with $P_1, ...P_V$ as the `bra's and $\lambda_1, 
...\lambda_V$ as the coordinate `ket's. In particular,
\be
\nu (\lambda ) =\int d\lambda_2 ...d\lambda_N \,\,\nu(\lambda, 
\lambda_2,...\lambda_V) = \sum_{i=1}^V |P_i (\lambda)|^2
\label{mic4}
\ee
for the eigenvalue distribution at finite $V$ and $N_f$. Eq.~(\ref{mic2}) 
summarizes all eigenvalue correlations.

\begin{figure}[htbp]
\centerline{\epsfysize=55mm \epsfbox{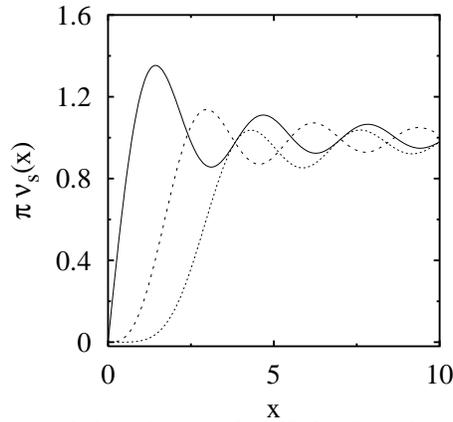}}
\caption{Microscopic spectral distribution for ChGUE with $N_f=0$ (solid),
$N_f=1$ (dashed), $N_f=2$ (dotted).}
\label{fig-mic}
\end{figure}

In the microscopic regime $m<1/V$. Consider the simplest case where $m=0$,
for which (\ref{mic3}) is just the condition for the generalized 
Laguerre-Selin polynomials
\cite{VERBAR}. In the microscopic regime $V\rightarrow\infty$ with 
$V\lambda$ fixed, this distribution can be evaluated exactly in the form
\be
\nu_s (x) = (\Sigma^2 x/2) \left(J_{N_f}^2 (\Sigma x) -
J_{N_f+1} (\Sigma x) J_{N_f-1} (\Sigma x) \right)
\label{mic5}
\ee
in the sector with zero winding number. The dependence of $\nu_s (x)$ is
shown in Fig.~\ref{fig-mic}, for $N_f=0,1,2$. The larger $N_f$ the wider
the distribution at zero virtuality, following larger repulsion from the
fermion determinant. In our case $\Sigma=1$, but in general it is the
scale associated to $\la q^{\dagger}q\ra$ in the chiral limit as is
apparent from (\ref{gas3}).

In a finite box, $\la q^{\dagger}q\ra$  gets
renormalized by non-zero modes, and thereby develops a dependence on
$1/L$ along with $m$ (and of course $\Sigma$) as originally discussed by
Hansen and Leutwyler~\cite{HANSEN}. We will refer to it as $\Sigma (m,
L)$. Its dependence on the low energy parameters at the chiral point
such as the pion decay constant, scattering length etc. along with $1/L$
and can be organized using a double book-keeping in $\epsilon$ and $p$.

The correlations between eigenvalues in the microscopic limit can be 
constructed using similar arguments. In particular, the connected 
correlator between
two eigenvalues $x$ and $x'$ in the microscopic regime for $n=0$ is
\cite{VERBAR}
\be
\label{mic6}
&&\nu_s (x, x') = \Sigma^2 xx' \\
&&\left(\frac
{xJ_{N_f} (\Sigma x)J_{N_f-1} (\Sigma x')
-x'J_{N_f} (\Sigma x')J_{N_f-1} (\Sigma x)}{x^2-{x'}^2}\right)^2\,.\nonumber
\ee
The connected two-point correlator (\ref{mic6}) obeys the consistency condition
\be
\int dx\  \nu_s (x, x') =0
\label{cons}
\ee
which reflects on the conservation of the total number of eigenvalues.
We will see below that the class of all n-point unconnected correlators 
may be resummed to give the microscopic eigenvalue distributions with 
$m\neq 0$ in the double scaling limit. Eq.~(\ref{mic6}) can be used to check
the off-diagonal sum rules discussed by Leutwyler and Smilga~\cite{SMILGA}.
The microscopic two-point correlator (\ref{mic6}) may be used to address
some issues related to the flavor admixture and spectral rigidity in the 
microscopic limit, as we now discuss.
\begin{figure}[htbp]
\centerline{\epsfysize=55mm \epsfbox{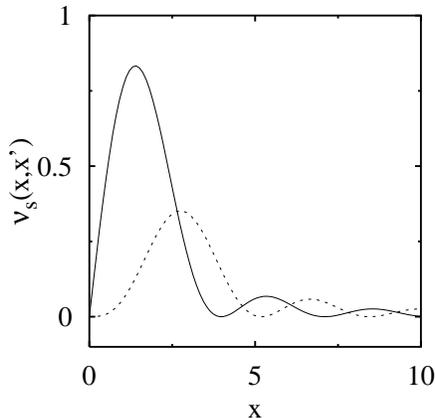}}
\caption{Microscopic connected correlator for ChGUE $N_f=0$ (solid)
and $N_f=1$ (dashed line) with an arbitrary normalization.}
\label{fig-corr}
\end{figure}

As we have stressed above, the universality of (\ref{mic5}) irrespective
of the gaussian measure adopted is guaranteed by {\bf i)-iii)} as discussed in
section {\bf III}, thanks to the modified power counting. These
observations are of course in agreement with those made by Nishigaki and
Damgaard~\cite{NISHIGAKI} in the context of polynomial weights, though
more general, as they apply to a field theory. Although
the present arguments apply to QCD in even-dimensions they can be easily
modified to accommodate QCD in odd dimensions as discussed by
Verbaarschot and Zahed~\cite{ODD}. This point is noteworthy as it
implies that the issue of the instantons and a vacuum angle
 is not crucial for the present
observations.

\subsection{Flavor Mixing}

It was noted early on by Nowak, Verbaarschot and Zahed~\cite{NOWAK}, 
that the eigenvalue correlator in the macroscopic limit reflects on
important aspects of the QCD vacuum as a disordered medium. In particular,
it was suggested using the instanton liquid model that 
the admixture of flavor (strangeness)
 is suppressed in the ground state (average of
all n-states) if the instanton density is closer to the metallic regime.
A measure of this suppression is given by the two-point correlator in the
eigenvalue distribution, and could be directly related to the level
variance in the macroscopic limit. Zweig's rule is supported by Wigner-Dyson
statistics in the macroscopic limit.

In light of our present discussion, 
similar questions may be asked in the microscopic limit, for which it is
easy to show that the Zweig's violating correlator investigated in
\cite{NOWAK}, that is $\la s^{\dagger}su^{\dagger}u\ra$ can be written 
in terms of~(\ref{mic6}) in the $n=0$ state with $y_i=m_iV$ as,
\be
i^2\la s^{\dagger}su^{\dagger}u\ra_0 (y_u, y_s ) =
\int\!dx dx'\frac {y_u}{x^2\!+\!{y_u}^2}
	\frac {y_s}{x'^2\!+\!{y_s}^2} \nu_s (x, x').
\label{mic8}
\ee
In particular,
\be
\displaystyle\lim_{y\to 0}
{{i^2\la s^{\dagger}su^{\dagger}u\ra}_0\over{y_u y_s}}
= \frac{\Sigma^4}{16N_f^2 (N_f+1)}
\label{mic9}
\ee
which is seen to break the cluster decomposition. The admixture
of flavor in the microscopic limit in an n-state is universally
given by the multipoint correlators in the microscopic limit.

\subsection{Spectral Rigidity}

The microscopic spectral density $\nu_s (x, x')$ characterizes
the level spacing distribution a quantum leap apart, not only 
at zero virtuality but also in the bulk of the spectrum. Indeed,
it carries important information on the spectral statistics in the 
microscopic regime, such as the spacing distribution or the level 
variance. It also plays 
a major role at the interface of the theory of disordered systems and
the semiclassical theory of quantum chaos, as we will try to elaborate 
further on in the end.

To exemplify some of these points, let us consider in more detail the 
genealogy of $\Sigma_2$, the variance in the number of single particle 
levels within an energy interval $\Delta\lambda$ centered around 
$\lambda_0$, in the microscopic regime 
$\Delta x=V\Delta(\lambda-\lambda_0)\gg1$  but
fixed as $V\rightarrow\infty$. This regime is characterized by constant
mean density, $\nu_s(x)=\Sigma/\pi$,\footnote{Actually this assumption
is not really required, as unfolding procedures of spectra may be
used~\cite{MEHTA}.} and in the decomposition of the
full two point correlator,
 $\nu_s (x_1,x_2) =\nu_s (x_1)\nu_s(x_2)[ \delta (x_1-x_2) -R(x_1,x_2)]$,
the cluster function $R(x_1,x_2)$ is given as
\be
R(s,\Lambda) = \left[\f{\Lambda\sin{\Sigma s}+
	(-1)^{N_f}s \cos{2\Sigma\Lambda}}{\Lambda s\pi} \right]^2
\ee
with $\Lambda=(x_1+x_2)/2$ and $s\equiv x_1-x_2$. In the bulk 
($\lambda_0\neq0$) the second term may be neglected and the cluster function
depends only on $s$~\cite{DYSON}. The edge ($\lambda_0\!=0\!$) is 
special
due to the vanishing of the density of states as required by chiral
symmetry. The level number variance, $\Sigma_2(N)$ is expressed 
through the cluster function $R$ as
\be
\label{sig2}
  \Sigma_2(N) = \int_0^N\!dx_1 dx_2 R(s,\Lambda)\,.
\ee
Specifically
\be
\Sigma_2 (N) = \frac 2{\beta\pi^2} \log{N}
\label{sta3}
\ee
with $\beta=2$ in the bulk and twiced at the edge, $\lambda_0 =0$. For the
orthogonal and symplectic case $\beta=1$ and $4$ 
respectively. Both for the quenched and unquenched cases, the spectral 
rigidity is $2{\rm ln}N/(\beta\pi^2)$ as opposed to $N$ for Poisson
statistics. The decrease in the variance is an indication of a more 
rigid spectrum due to level repulsion~\cite{WIGNER}.

\subsection{Spacing Distribution}

The probability $P(s)$ to find two energy levels in the quark spectrum a 
distance $s$ apart in the microscopic limit, that is 
$s=(\lambda_i-\lambda_j)V/\Delta$ fixed as $V\rightarrow\infty$, is also
intimately related to the properties of the two-point correlator (\ref{mic6}).
For $s\ll1$, that is as $\lambda_i\rightarrow \lambda_j$, we immediately see
from the Vandermonde determinant (\ref{van}) that $P(s)\sim s^{\beta}$ with
$\beta=2$. This is the usual level repulsion between two neighboring levels as 
predicted by Wigner~\cite{WIGNER}. It follows from the random lore for
$2\times 2$ matrices, since in this limit the rest of the eigenvalue spectrum 
decouples. Clearly it is independent of the number of flavors, except when
one of the level is at zero virtuality. In this case, further repulsion is 
introduced by the fermion determinant and the winding number $n$ in the 
massless case as is evident from Fig.~\ref{fig-corr}.

For $s\gg1$ (still in the microscopic limit), the spacing distribution
is solely  
governed by the asymptotics of the two-point correlator (\ref{mic6}). For
$R(s) \sim 1/{\beta s^2}$, the level distribution asymptotes
$-{\rm ln} P(s) \sim \beta s^2$ as was shown by Dyson~\cite{DYSON} using the
the Coulomb gas treatment for (\ref{gas3}) in the quenched case. In the 
unquenched case, the asymptotics of the two-point correlator is unaffected
by the fermion determinant, resulting into the same level distribution. 
Indeed, for any $N_f$, it is  
straightforward to see that the fermion determinant contributes to the
potential term (one-body) not the interaction term (two-body). Since it
is the latter that conditions the level spacing distribution, the 
self-quenching of this result is immediate. Hence, the whole level spacing 
distribution for (\ref{gas3}) could be described by the standard Wigner 
surmise~\cite{WIGNER}
\be
P(s) = A s^{\beta} e^{-B \beta s^2}
\label{surm}
\ee
where $\beta=2$ and $A,B$ are pure numbers. Which is drastically different
from the Poisson distribution $P(s)=e^{-s}$, for uncorrelated states.

\subsection{Relevance to Lattice}

By now the present observations have been generalized to the case of finite
winding number, as well as the two other Wigner ensembles: ChGOE and ChGSpE
by Verbaarschot~\cite{MANY1}, and  checked against
lattice Dirac spectra by Sch\"afer, Weidenm\"uller and Wettig~\cite{MANY2},
and others~\cite{OTHERS}. In regard to the lattice measurements, we would 
like to point at few issues in regard to the strong coupling aspects of the
simulation as well as the approach to the weak coupling regime. Some of our
remarks will also relate to certain aspects of the quenched approximation.

In strong coupling, the Kogut-Susskind action is characterized by an exact
$U(1)\times U(1)$ symmetry. The analogue chiral random matrix model is ChGSpe
\cite{MANY1}. So the strong coupling regime is characterized by universal
microscopic spectral oscillations, even
though Lorentz and full chiral symmetry
are still absent. We also expect quantitative changes in the sum rules for
quenched and unquenched simulations since the character of the coset and hence
the number of Goldstone mode changes in this case~\cite{MOREL}. 

In weak coupling, the Kogut-Susskind action is expected to undergo a 
transition to a phase with continuum Lorentz and full symmetry. A structural
change from ChGSpe to ChGUE is therefore expected, for otherwise the true
continuum would not have been reached. This point can be checked
at the level of the microscopic sum rules, with due attention to scale
translations, {\it e.g.} $V, m$ (see below).

In symmetric boxes an accurate measurement of the finite volume chiral 
condensate in weak coupling can be reached by checking the microscopic 
sum rules, or simply monitoring the Bessel oscillations. This
means an accurate assessment of the chiral condensate at order $p^2$, as
well as the pion decay constant to the same order
through possible finite volume effects.

In asymmetric boxes, the chiral condensate gets modified by temperature
effects which can be assessed using a double expansion in $\epsilon$ and $p$ (see 
below) at low temperature (small asymmetry). At high temperature, the chiral
condensates disappears as the level spacing near zero becomes larger than 
$1/V$. For temperatures $T\sim T_c$, universality arguments may be used to
argue for critical sum rules~\cite{USCRITICAL}. For $T>3T_c$, the virtual 
spectrum is gapped by the lowest Matsubara mode $\lambda\sim \pi T$, 
except for thermodynamically irrelevant zero modes.

Finally, the relevance of the present concepts to lattice QCD in the strong
coupling indicates that they are also applicable to lattice electronic systems
with ferromagnetic or antiferromagnetic ground-state order. An example is the
undopped Lanthanum-Copper-Oxide material for high temperature superconductors
\cite{HIGHTC}.

\section{Double Scaling Regime}

\subsection{The Formulae}

In the transition regime $mV\sim 1$ the current quark masses affect the 
distribution of quark eigenvalues quantitatively for $N_f\neq 0$, through
the occurrence of the fermion determinant. Surprisingly enough, the non-local
character of the latter supersedes the effects of local weights to upset
the conventional universality arguments based on polynomial weights in the
microscopic limit. This point was originally shown by
Jurkiewicz, Nowak and Zahed~\cite{JUREK} using the supersymmetric method
for $N_f=1$ and zero winding number, and more recently by Damgaard and 
Nishigaki~\cite{DAMGAARD} and Wilke, Guhr and Wettig~\cite{WILKE} using
the orthogonal polynomial method with arbitrary $N_f$ and winding number $n$
\footnote{Incidentally, this construction applies to any additive part one 
adds to the fermion determinant say $T$ provided that it is suitably rescaled 
$T\rightarrow VT$ fixed, in the microscopic regime.}.

The massive spectral density in the double scaling limit yields by 
definition the quark condensate in a fixed winding number configuration 
$n$, but with different $m_s$ sea and $m_v$ valence quark masses in
general. Specifically, 
\be
i\la q^{\dagger} q \ra_{n} (\xi , y) = 
\int_0^\infty\!\! dx\, \frac {2\xi}{x^2+\xi^2} \nu_{s,n} (x , y)
=\xi \Sigma_n^2 (\xi , y) 
\label{doux}
\ee
with the rescaled valence quark mass $\xi=\Sigma m_vV$ and a finite sea
quark mass $y=\Sigma m_sV$. We are using different notations for $\xi ,
y$ to highlight their different origins. For $\xi = y$ the result is
just the massive quark condensate in a winding number configuration
$n$. The physical quark condensate with $\xi =y$ follows by suitably
averaging over all windings $n$.
For $N_f=1$ and $n=0$ the
massive spectral density is ($x=\Sigma V\lambda$)
\be
\frac 1{\Sigma^2} \nu_{s,0} (x, y) =&&  \frac x2 \left[J_0^2(x)+J_1^2(x)\right]
\label{dou1}\\
&&\hspace*{-10mm}-\frac {J_0(x)}{I_0(y)} \frac x{x^2+y^2} 
	\left[y J_0(x)I_1(y) + x J_1(x) I_0(y)\right]
	\nonumber
\ee
for $x>0$. For $y=0$ (\ref{dou1}) reduces to (\ref{mic5}) with $N_f=1$.
\begin{figure}[htbp]
\centerline{\epsfysize=42mm \epsfbox{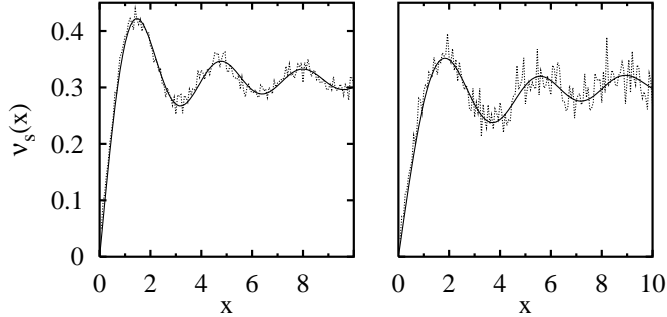}}
\caption{Massive microscopical spectral density from an ensemble of
50000 N=50 random matrices (dotted line) for $m_s=1$ (left) and
$m_s=0.1$ (right), $N_f=1$, and the analytical
formula~(\protect\ref{dou1}) (solid line).}
\label{fig-micdens}
\end{figure}

\subsection{Sum Rules}

The microscopic spectral density (\ref{dou1}) yields new microscopic
sum rules with sea quark effects included. These sum rules are not a priori
amenable to the finite volume partition function discussed above simply
because (\ref{dou1}) selects solely the sea quark mass effect, which is
not a physical concept. Finite volume partition functions for QCD with 
different sea and valence quark masses can of course be simply constructed 
by using quark ghost fields~\cite{MOREL}. Such discriminations are easily
achievable on the lattice, and hence are of theoretical relevance. An 
example being the detailed valence quark study undertaken recently by 
Chandrasekharan and Christ~\cite{COLUMBIA}, for a fixed sea quark mass 
contribution.

The derivative of (\ref{doux}) with respect to the valence quark mass
generates a string of sum rules that are now sensitive to the effects 
of the sea quark mass $y$. Specifically, for $n=0$, $N_f=1$, and fixed
$\xi , y$
\be
\frac 1{V^2}
\langle\langle\sum_k'\frac 2{\lambda_k^2 +m_v^2}\rangle\rangle_0 = 
\Sigma_0^2 (\xi , y)
\label{sminew}
\ee
where the double averaging on the left-hand side involves 
$m^{|n|} \prod'_k (\lambda_k^2 + m_s^2)$ as part of the measure,
in contrast to the one used in (\ref{smi4}) with $m_s=0$. 
$\Sigma_0 (\xi , y)$ depends on the rescaled valence and sea quark
masses independently. Its analytical form is 
\be
\frac{\Sigma_0^2 (\xi , y)}{\Sigma^2} &=&
\label{long}
I_0(\xi ) K_0 (\xi ) + I_1 (\xi ) K_1 (\xi) \\
&+&\f{2 K_0(\xi)}{y^2-\xi^2} \left(
	\xi I_1(\xi)-\f{y I_1(y)}{I_0(y)} I_0(\xi) \right)\,. \nonumber
\ee
For $y=0$ it is in agreement with the result discussed  by 
Verbaarschot~\cite{VERLET}. In the limit of equal sea and valence quark
masses the expression simplifies to
\be
\frac{\Sigma_0^2 (\xi,y=\xi)}{\Sigma^2} &=&
	\f{I_1(\xi)}{\xi I_0(\xi)} \,.
\ee

\begin{figure}[htbp]
\centerline{\epsfysize=45mm \epsfbox{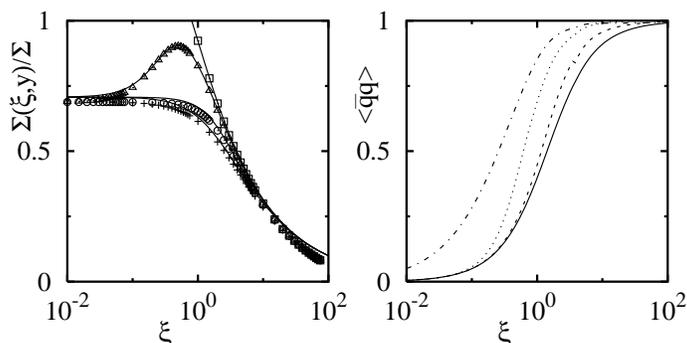}}
\caption{{\bf a} Sum rule~(\protect\ref{sminew}) obtained numerically
from an ensemble of 5000 $N=50$ random matrices and the analytical result.
{\bf b} The quark condensate for various masses. See text.}
\label{fig-doub}
\end{figure}

In Fig.~\ref{fig-doub}a we show the behavior of $\Sigma_0(\xi,y)/\Sigma$
for $y=0$ (plus signs), $y=\xi$ (crosses), $y=5\xi$ (triangles) and for
fixed $m_s=1$, that is $y=1/N$ (boxes) from numerical simulation of 5000
 $N=50$ random matrices and the corresponding analytical
form~(\ref{long}) (lines).
For fixed and finite sea quark mass Eq.~(\ref{sminew}) diverges 
 for small $\xi$ as $-\log{\xi}$. 
In Fig.~\ref{fig-doub}b we show the behavior of 
the $n=0$ condensate from~(\ref{mic5}) for the same set 
of masses, namely $y=0$ (solid line), $y=\xi$ (dashed line), $y=5\xi$
(dotted line) and fixed $m_s=1$ (dashed-dotted line) . 


For two massive sea quarks the result for the microscopic spectral density
was worked out by Damgaard and Nishigaki~\cite{DAMGAARD}, and Wilke, Guhr and 
Wettig~\cite{WILKE} for arbitrary $n$. In this case, the analogue of
(\ref{long}) is 
\be
\label{nf2}
  &&\frac{\Sigma_0^2(\xi,y_1,y_2)}{\Sigma^2} = I_0K_0+I_1K_1- \\
  &&\frac{y_1^2\!-\!y_2^2}{y_1I_1^aI_0^b\!-\!y_2I_0^aI_1^b}
  \frac 2{y_1^4(y_2^2\!-\!\xi^2)\!+\!y_2^4(\xi^2\!-\!y_1^2)\!+\!
	\xi^4(y_1^2\!-\!y_2^2)}
  \times \nonumber \\
  &&\Big\{ y_1y_2I_1^aI_1^b \left[(y_1^2\!-\!y_2^2)I_0K_0\!+\!
	(y_2^2\!-\!\xi^2)I_0^aK_0^a\!+\!(\xi^2\!-\!y_1^2)I_0^bK_0^b\right]
  \nonumber\\
  &&+\!(y_1I_1^aI_0^b\!+\!y_2I_1^bI_0^a) 
	\left[\xi(y_1^2\!-\!y_2^2)I_0K_1\!+\!y_1(y_2^2\!-\!\xi^2)I_0^aK_1^a
	\right. \nonumber \\
  &&\left.+\!y_2(\xi^2\!-\!y_1^2)I_0^bK_1^b\right] -I_0^aI_0^b 
	\times \nonumber \\
  &&\left[\xi^2(y_1^2\!-\!y_2^2)I_1K_1\!+\!
	y_1^2(y_2^2\!-\!\xi^2)I_1^aK_1^a\!+\!
	y_2^2(\xi^2\!-\!y_1^2)I_1^bK_1^b\right] 
	\Big\} \nonumber
\ee
with $I_i=I_i(\xi), I_i^a=I_i(y_1)$ and $I_i^b=I_i(y_2)$, respectively
and similarly for $K$'s.
For one massive quark and $N_f-1$ massless quarks, the sum rule is given
by
\be
 &&\frac{\Sigma_0^2(\xi,y)}{\Sigma^2} = I_{N_{\!f}}K_{N_{\!f}}+\inm\knp \\
 &&-\frac{y^2}{N_f (y^2-\xi^2)} 
	\Big[\inm\knp-\inm(y)\knp(y)\Big] \nonumber \\
 &&+\frac{y^2\inp(y)}{N_f\inm(y)(y^2\!-\!\xi^2)} 
	\Big[\inm\knm\!-\!\inm(y)\knm(y)\Big]\,. \nonumber
\ee
For one massive quark and $N_f-1$ massless quarks, the sum rule for the
massive quark is $\Sigma^2 (y, y) = \Sigma^2 I_{N_f}/2yI_{N_f-1}$ for
$n=0$ as also noted by Damgaard~\cite{DAMGAARD}. We note that in this
case $Z=Z_0$, which is the partition function in the $n=0$ sector. In
particular, $i\la q^{\dagger} q\ra = i\la q^{\dagger} q\ra_0$. 
Additional sum rules involving diagonal and off-diagonal eigenvalue 
correlations are of course possible. The off-diagonal ones will involve 
the n-point correlation functions of eigenvalues with massive quarks.

\begin{figure}[htbp]
\centerline{\epsfysize=45mm \epsfbox{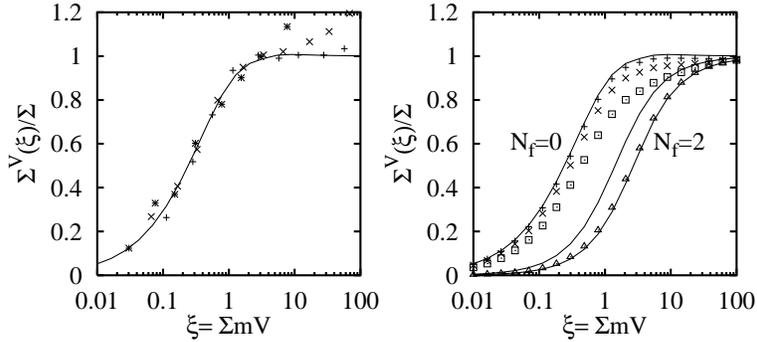}}
\caption{Normalized condensate. See text.}
\label{fig-columb}
\end{figure}
Recently Chandrasekharan and Christ~\cite{COLUMBIA}
have analyzed in details the behavior of the valence 
quark condensate for fixed sea quark mass $m_sa=0.01$,
for staggered $N_f=2$ QCD, over several decades of the 
valence quark mass in an asymmetric lattice 
$V=4\times 16^3 a^4$. Their analysis was carried for various
lattice coupling $\beta=6/g^2$ (varying lattice spacing $a$), 
around the chiral transition point $\beta_c=5.275$. Their
underlying quark spectrum is clearly sensitive to temperature, 
and an analysis of this point using a macroscopic model distribution 
was discussed by Nowak, Papp and Zahed~\cite{NPZT}. 
For small asymmetries, $\beta<\beta_c$,
we may still assume that the conditions {\bf i)}-{\bf iii)} 
summarized at the end of section {\bf IIIA} are still valid, 
in which case $i\la q^{\dagger} q\ra_0 (\xi , y)$ may be 
amenable  to a lattice comparison. For that, we need to identify
$\Sigma_La^3$ for different $\beta$ and the proper dimensionless
combination $\xi_L= m\Sigma_L V$ on the lattice. For the former,
we have: $\Sigma_L a^3 = 0.2217$ ($\beta=5.245$), 
$\Sigma_L a^3=0.1357$ ($\beta =5.265$) and $\Sigma_L a^3 =0.0543$
($\beta =5.270$). 
 
In the left part of Fig.~\ref{fig-columb}
we show the lattice results for the normalized condensate
versus the rescaled valence quark mass, $\xi_L$.
The line is the zero flavor result~\cite{VERLET}
in the $n=0$ topological sector seemingly in good agreement with the two
flavor lattice data (symbols for different $\beta$'s).
In the right we show the two flavor result~(\ref{nf2}) for equal valence
masses, $y=m_v\Sigma V$. The solid lines represent $N_f=0,1,2$ massless
flavors~\cite{VERLET}, respectively, the plus signs are for the lattice
data with $\beta=5.245$ while the crosses for $\beta=5.270$. We conclude
that the used sea mass on the lattice is still too large making
the simulation analogous to a quenched one ($N_f=0$). Decreasing
the sea mass by a factor of ten we start to see the deviation from
the quenched result (boxes), and lowering the mass by a factor of
hundred we recover the $N_f=2$ zero mass limit (triangles). The same
effect can be achieved dialing $\beta$ closer to the chiral transition
point, where the condensate, $\Sigma$,  disappears. This tendency is
seen in the right part Fig.~\ref{fig-columb} for $\beta=5.270$.

For small valence quark mass $\xi<0.1$ limit of Eq.~(\ref{long}), we find
\be
\frac{\Sigma_0^2 (\xi , y)}{\Sigma^2} =
	-\left(\log{\f{\xi}2}+{\bf C}\right) \left(
	1-\f{2I_1(y)}{yI_0(y)}\right)+\f 12 \,.
\label{unqsp}
\ee
In the large sea mass limit we recover the quenched limit, 
 $-\log{\xi/2}-{\bf C}+1/2$ with a logarithmic divergence.

\subsection{Universality}

The result (\ref{dou1}) is universal for the QCD spectrum as we now show. 
Following Smilga and Stern~\cite{SMILGASTERN}, we can rewrite the massive 
fermion determinant for a fixed background as ($n=0$)
\be
\Delta_m =\prod_{\lambda_k >0}(\lambda_k^2 + m^2)^{N_f}
=\Delta_0 {\rm exp}\left(N_f\sum_{\lambda_k} {\rm ln} (1+\frac 
{m^2}{\lambda_k^2})\right) \,.
\label{dou1a}
\ee
Hence, the density of eigenvalues for finite $m$ reads
\be
\label{dou1b}
\rho (\lambda , m ) =
 \frac 1V \langle {\rm exp}\!\left(N_f\!\!\int_0^{\infty}\!\!\!\!\!d\lambda'
 \ \omega (\lambda'\!, A)\ {\rm ln} (1+\frac {m^2}{\lambda'^2})\!\right)
\omega (\lambda, A) \rangle_A \,. \nonumber
\ee
where the averaging is over $A$ including $\Delta_0$, the fermion
determinant with zero mass quarks, and $\omega (\lambda, A)$ is the
unaveraged spectral operator
\be
\omega (\lambda , A) =\sum_k \delta (\lambda -\lambda_k [A])\,.
\label{dou1c}
\ee
In terms of the rescaled variables $x=V\lambda$ and $y=Vm$,
the expression (\ref{dou1b}) reads
\be
\label{dou1d}
\nu_s (x,y) =
\Big\langle {\rm exp}\left(N_f\!\!\int_0^{\infty}\!\!\!\!dx'\ 
  \omega (x'\!, A)\ {\rm ln} (1\!+\!\frac {y^2}{x'^2})\right)
\omega (x, A) \Big\rangle_A \nonumber
\ee
which involves all the moments of the microscopic spectral
distribution. Specifically,
\be
\label{dou1e}
\nu_s (x,y) =\nu_s (x) + 
N_f\!\!\int_0^{\infty}\!\!\!\!{dx'}\ {\rm ln} 
	\left( 1\!+\!\frac {y^2}{{x'}^2}\right) 
\langle \omega (x'\!, A)\ \omega (x, A) \rangle_A + ...\nonumber
\ee
which is the weighted density-density correlator (second moment)
in the microscopic limit. Each of the expectation value is carried
in the n=0 state with massless quarks.
All the moments are universal and given 
by random matrix theory. For $N_f=1$ the result for (\ref{dou1d})
is (\ref{dou1}).

\subsection {Range of validity}

Since $mV\sim 1$ is the transition region in which the quark condensate 
(\ref{gas4}) is no longer averaging to zero, the pionic zero 
modes in (\ref{gas2}) are no longer dominant. What is the 
range of validity of (\ref{dou1}) in QCD?
The answer follows by noticing that (\ref{gas1}) can be extended to one
extra dimension, that is a Lagrangian in 1+4 dimensions. The support on the 
fifth direction is $[0,\beta]$ with periodic boundary condition for 
$U(x_5+\beta, x)= U(x_5,x)$. Here $\beta$ plays the role of a `temperature' and  
$-{\rm ln} Z/\beta$ for large $\beta$ is just the ground state 
energy of the singlet Hamiltonian described by
\be
{\bf H}_{1+4} = \frac {\vec{\bf L}^2}{2V} - \frac {mV}2 {\rm Tr} (U + U^{\dagger})
\label{lpt}
\ee
where $\vec{\bf L}^2$ is the Laplace-Beltrami operator on the $SU(N_f)$ 
manifold. Note that the present construction is analogous to the one 
discussed by Leutwyler~\cite{XLEUTWYLER} in $1+3$ dimensions. 

\begin{figure}[htbp]
\centerline{\epsfysize=35mm \epsfbox{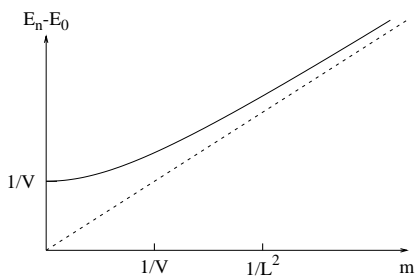}}
\caption{Mass gap versus ${m}$ from (\ref{lpt}).}
\label{fig-gap}
\end{figure}

For $mV=0$ the spectrum is that of a spherical top with the irreducible
representations of $SU(N_f)$ as eigenfunctions. The spectrum is 
$1/V$ spaced with a mass gap. The ground state wavefunction is a constant on 
$SU(N_f)$. The first excited state is $N_f^2$ degenerate. For
$mV\leq 1$ the degeneracy is lifted. For $N_f=2$, the degeneracy is 4 with a 
triplet (pions) and a singlet (sigma) state. With increasing $mV$, 
the triplet states are pushed down and the singlet state up. The former
merge into the physical pion mass as shown schematically in 
Fig.~\ref{fig-gap}. We note that $m<1/V$ puts the mass range into
the ergodic regime, and $1/V<m<1/\sqrt{V}$ to the diffusive 
regime~\cite{PRLUS}. The transition to the macroscopic regime with a  
mass gap given by $\sqrt{m}$ (remember that $m_{\pi}^2 =2m$ with our 
conventions) sets in for $mV \sim 10$.

\subsection{New Results}

Recently Damgaard~\cite{DAMGAARD} has suggested an interesting relation
for $\nu_s (x,y)$ in terms of the finite volume partition function involving
$1+2$ flavors with respective masses $(y, ix,ix)$, using a property of the 
chiral random matrix measure. Generically, the microscopic spectral density
reads~\cite{DAMGAARD}
\eq
\label{e.damgaard}
\rho(\lm)=C\cdot|\lm|\prod_i
(\lm^2+m_i^2)\f{Z(m_1,\ldots,m_{N_f},i\lm,i\lm)}{Z(m_1,\ldots,m_{N_f})}
\,.
\eqx
Schematically it means that
\eq
\cor{\dl(z-\lm)}=\cor{\prod (\lm_i-z)^2}\cdot |\lm|\prod_i (\lm^2+m_i^2)
\eqx
i.e. we may trade in a Dirac delta function for two {\em additional}
flavors with imaginary masses. The origin of this formula in the
random matrix model is very simple. One has just to insert the Dirac
delta function into the expression of the RMM partition function
written in terms of an integral over eigenvalues. Then the two
flavors and the factors outside the expectation value just follow
from contracting with the Vandermonde determinant. The term with the
probability distribution $e^{-NV(M)}$ does not contribute in the
microscopic limit. 

One would like to understand this formula from the point of view of QCD.
The main difficulty is that in this regime of QCD, we are dealing only
with the Goldstone boson degrees of freedom and all the explicit
dependence on the spectrum of the Dirac operator seems to have been
lost. In particular the low lying states are composite states in terms
of the quark fields.

The main unsolved problem is to find a nonlocal operator in the mesonic
picture whose expectation value would correspond to the microscopic
spectral density~\cite{NEWPOUL}.

%

\section{Macroscopic Regime}

In the macroscopic regime $mV>1$, the constant modes in (\ref{gas2}) 
cease to have a preferential role, and we are back to full QCD
with (\ref{gas1}), (\ref{gas5}) (or even other unexplored power
countings) as starting points in power counting.

However, there are a number of problems in QCD where power counting
(in the sense exposed) breaks down. Examples are the strong CP 
problem, the \uone problem, and phase transitions to cite afew. 
Since these problems involve in an intricate way the issues of large
volumes, small quark masses, zero modes {\it etc.} it is useful to
address in a framework where these quantities are simply separated
in a way that allows for an analytical treatment. More importantly, 
the framework should be able to include some generic dynamical aspects
of QCD in the form of few vacuum moments and symmetry. 

The chiral random matrix models introduced above for studying the
microscopic character of Dirac spectra, offer such an example. In
many ways they can be regarded as a schematic description of the 
chiral physics at work in a cooled lattice QCD configuration within
a finite Euclidean box $V$. Since these models allow for closed form
results in terms of $m$, $V$, $N_f$,  {\it etc.}, they are very useful
for addressing some subtle aspects of the thermodynamical limit in the
presence of zero or near-zero modes whether in vacuum or matter.

\subsection{Wide Correlators}

Chiral correlations in QCD involve usually eigenvalues which
are separated by a macroscopic distance in the Dirac spectrum.
They are defined for $V\rightarrow\infty$ and fixed $\lambda_i-\lambda_j$. 
Correlations over these distances will be discussed in this
part of the lecture. For bounded spectra, Ambj\o rn,
Jurkiewicz and Makeenko~\cite{AMBJORN} have shown that the smoothened
n-point correlation functions could be classified by the support
of the spectral densities, independently of the specifics of the 
random ensemble and genera of the topological expansion. 
In QCD the Dirac spectrum is not bounded due to a strong ultraviolet
tail $\rho (\lambda) \sim |\lambda|^3$. This tail, however, is not very 
important for infrared physics, except for multiplicative renormalization
factors and anomalies. In a lattice formulation, this tail can be subtracted 
through a cooling procedure. The resulting spectrum is likely bounded,
(except for possible tails which may be triggered by a partial loss
of confinement), in which case the results discussed by Ambj\o rn, Jurkiewicz
and Makeenko~\cite{AMBJORN} for the hermitian case, and Janik, Nowak, Papp and 
Zahed~\cite{USBIG} for the non-hermitian case may apply. An early account
may be found in the analysis by Nowak, Verbaarschot and Zahed of the instanton 
liquid model~\cite{NOWAK}.

\subsection{\uone problem}

An important problem in QCD relates to the fact that the $\eta'$ in
nature is much more massive than the $\pi , K, \eta$ system. It is
believed that the discrepancy in mass is related to the fact that
the \uone current in QCD is anomalous. Indeed, using Ward identities
it follows that (Minkowski space)~\cite{CREWTHER}
\be
i\chi_{\rm top} = -\frac{im}{N_f^2}\langle\overline{q}q\rangle
+\frac {m^2}{N_f^2} \int d^4x \langle T^* \eta_0 (x) \eta_0 (0) \rangle
\label{on1}
\ee
where the topological susceptibility reads 
\be
\chi_{\rm top} = \int d^4x \langle T^* \Xi (x) \Xi (0) \rangle
\label{on2}
\ee
and the topological density is $8\pi\Xi (x)= {\alpha_s}E^a\cdot B^a (x)$.
The singlet current is $\eta_0 (x) =\overline{q}i\gamma_5 q (x)$.
For small $m$ a gap in the singlet correlator requires that $\chi_{\rm top}\neq 
0$. This is the presently accepted view for the resolution of the \uone 
problem, although some difficulties may be noted~\cite{YAZAUONE}.

The chiral random matrix models discussed in (\ref{rmm3}) offers a simple
way to model some of the aspects of the present problem. Indeed, the matrix
\be
\left(\begin{array}{cc}
ime^{i\theta}&A\\
A^{\dagger}&ime^{-i\theta}
\end{array}\right)
\label{un2}
\ee
with entries $n_L$ and $n_R$ can be viewed as a schematic description of
the topological zero modes in a finite volume $V$, with $\int \Xi =n_L-n_R$.
Since $\gamma_5$ is just
\be
\gamma_5 =
\left(\begin{array}{cc}
{\bf 1}&0\\
0&-{\bf 1}
\end{array}\right)
\label{un3}
\ee
it follows that the form of the anomaly in the present model is just
${\rm Tr}\gamma_5 = n_L-n_R$. The non-vanishing of (\ref{on2}) in 
QCD, will appear naturally in this model by summing over matrices
of the type (\ref{un2}) as in (\ref{rmm3}) with varying sizes. The
sizes will be gaussian distributed according to
\be
e^{-\frac{(n_L-n_R)^2}{2V\chi_*}}\,.
\label{un4}
\ee
Clearly $\langle (n_L-n_R)^2\rangle = V\chi_*$ with the determinant
set to $1$ in (\ref{rmm3}) (quenched case).

\begin{figure}[htbp]
\centerline{\epsfxsize=8.5cm \epsfbox{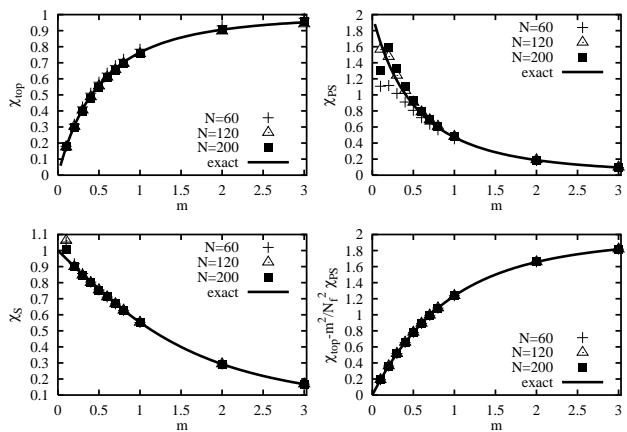}}
\caption{Normalized topological (upper left), pseudoscalar
(upper right), scalar (lower left) susceptibilities, and
Ward identity (lower right)  for $N_f=1$. The numerical simulations where
carried for fixed $2N=n_+ +n_-=  60, 120, 200$ on an ensemble of 
$10^4$, 5000 and 5000 matrices, respectively,
and $<(n_+-n_-)^2>=N\chi_{\star}=N$. The solid line is the analytical
result~\protect\cite{JANIKUONE}.}
\label{sus1fla}
\end{figure}

This model has been discussed recently by Janik, Nowak, Papp and Zahed
\cite{JANIKUONE}. Their results are shown in Fig.~\ref{sus1fla}, where
the topological susceptibility (~\ref{sus1fla}a), the pseudoscalar
susceptibility (~\ref{sus1fla}b) and the quark condensate
(~\ref{sus1fla}c) are studied versus $m$ for different values of
$N=n_L+n_R=V$. Clearly, the extrapolation to small values of $m$ warrant
larger and larger sizes $N$ (volume $V$) for the result of
Fig.~\ref{sus1fla}b to be meaningful. The solid curves are analytical
results in the thermodynamical limit. The numerical results were
obtained by sampling over matrices of different sizes using a Gaussian
distribution for the matrix elements as in (\ref{rmm3}) and the
distribution (\ref{un4}) for the size variations.

\subsection{CP problem}

The non-vanishing of the topological susceptibility means that the 
vacuum partition function of QCD depends on the value of the vacuum 
angle $\theta$. This point is also clear from our earlier arguments
as the free energy $F(\theta , m, V) =-{\rm ln}Z [\theta, m, V ]/V$  
was $\theta$ dependent by construction. QCD with at small $\theta$ implies
a free energy shift $F(\theta) -F(0) \sim {\chi_{\rm top}}\theta^2 /2$,
which violates $T$ and $P$. Since strong interactions are known empirically
to preserve $T$ and $P$ in the vacuum, this causes the strong CP problem.

A variety of scenarios have been put forward to resolve it ranging
from axions to confinement~\cite{AXICONFI}. This problem could have been
in principle settled on the lattice if it were not for the breakdown of
conventional Monte-Carlo algorithms at finite $\theta$. Recently,
Schierholz~\cite{SCHIERHOLTZ} has addressed this
issue in the context of the $CP^{n}$ model in two dimensions.
His conclusions that the model exhibits a first order phase transition at 
finite $\theta$ were challenged by Plefka and Samuel~\cite{SAM}. 

Our present framework offers a testing ground for the present
problem since both the issues of quenched, unquenched, finite size
and current quark effects can be dealt with explicitly. As was shown
by Janik, Nowak, Papp and Zahed~\cite{JANIKCP}
the model in many ways resemble the 
effective models discussed using effective Lagrangians in the saddle
point approximation. The advantage, however, is that the present model
allows for numerical simulations that test for the validity of such an
assumption, the subtlety of the chiral and thermodynamical limit and 
the importance of the numerical accuracy.
\begin{figure}[tbp]
\centerline{\epsfxsize=40mm \epsfbox{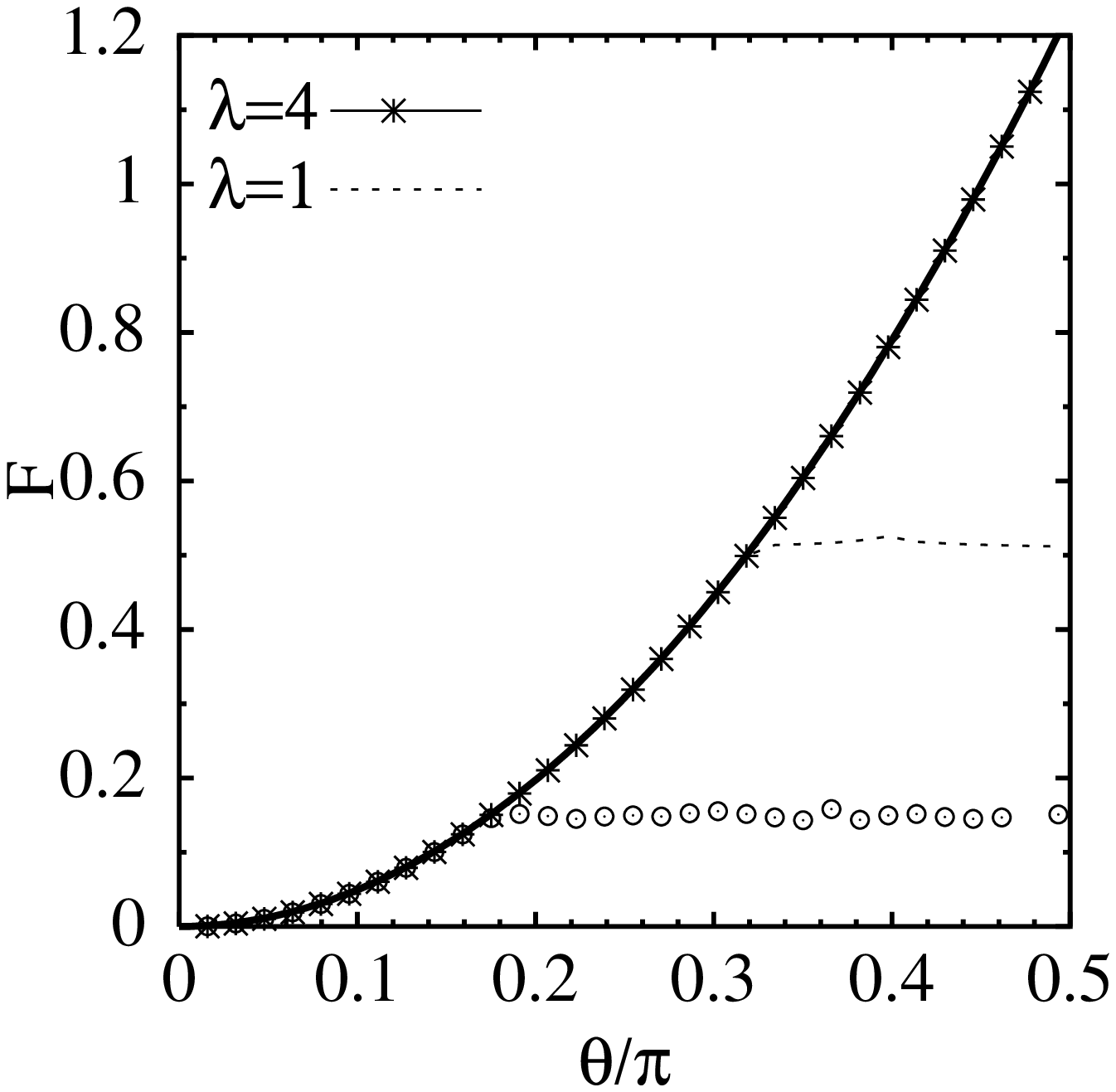} \hspace*{3mm}
 \epsfxsize=42mm \epsfbox{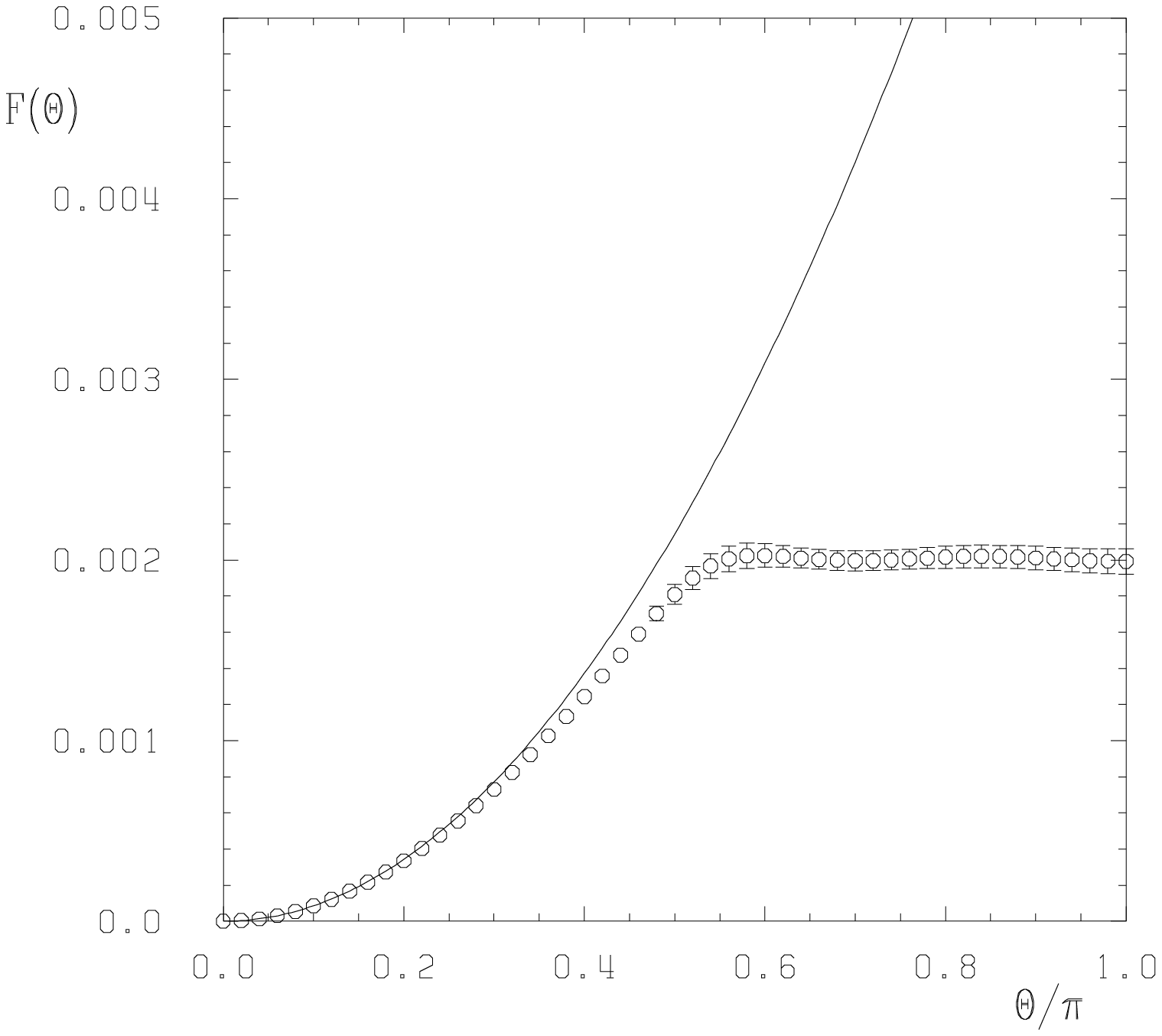}}
\caption{ChRMM (quenched) free energy for N=250 (left). The
normal precision calculation is shown by circles, the finite volume high
precision by dashed line and the large volume high precision by solid
line. Similar results are obtained for the free energy in the CP$^n$
model (right)~\protect\cite{SCHIERHOLTZ}.}
\label{fig-cpn}
\end{figure}

\section{Conclusions}

We have reviewed some of the motivations in the introduction of chiral
random matrix models to QCD problems, emphasizing some universal aspects
in the microscopic limit as well as some generic aspects in the
macroscopic limit. In particular, we have shown that by relaxing the GOR
relation new microscopic sum rules and random matrix models may be set up
in the context of QCD.

We have discussed  the double scaling regime and  we have presented
relations for the chiral condensate in a fixed winding number sector, 
that are sensitive to the sea and valence quark masses independently.

In the macroscopic regime, we have suggested that ChRMM although not universal,
exhibit generic aspects of the finite volume QCD problem that are useful
for addressing currently open problems in QCD, with embarrassing similarities
with bulk lattice simulations. ChRMM offer useful framework for
understanding the interplay between the thermodynamical limit, the
chiral limit and the precision of numerical algorithms.

Finally, we would like to conclude by pointing out that the chaotic aspect
of the quark eigenvalues near zero virtuality as first revealed by random 
matrix theory from the finite volume partition function, may be established 
directly from QCD and general principles without recourse to power 
counting~\cite{PRLUS}. These arguments carry to lattice formulations as well
as supersymmetric QCD.

\newpage
\section{Acknowledgments}
These notes are based on recent lectures and talks delivered 
by the authors at Kyoto (YKIS 1998), Trento (ECT$^*$ 1998), 
Cracow (Meson 98) and 
Zakopane (XXXVIII Course of Cracow School of Theoretical Physics). 
This work was supported in part by the US DOE grant DE-FG-88ER40388, by the 
Polish Government Project (KBN) grant 2P03B00814 and by the 
Hungarian grants FKFP-0126/1997 and OTKA-F019689.

\end{document}